# HOW PLANNERS DEAL WITH UNCOMFORTABLE KNOWLEDGE: THE DUBIOUS ETHICS OF THE AMERICAN PLANNING ASSOCIATION

By

Bent Flyvbjerg

Director, Professor, Chair

BT Centre for Major Programme Management

Saïd Business School, University of Oxford

Park End Street, Oxford OX1 1HP, UK

bent.flyvbjerg@sbs.ox.ac.uk

## ABSTRACT

With a point of departure in the concept "uncomfortable knowledge," this article presents a case study of how the American Planning Association (APA) deals with such knowledge. APA was found to actively suppress publicity of malpractice concerns and bad planning in order to sustain a boosterish image of planning. In the process, APA appeared to disregard and violate APA's own Code of Ethics. APA justified its actions with a need to protect APA members' interests, seen as preventing planning and planners from being presented in public in a bad light. The current article argues that it is in members' interest to have malpractice critiqued and reduced, and that this best happens by exposing malpractice, not by denying or diverting attention from it as APA did in this case. Professions, organizations, and societies that stifle critique tend to degenerate and become socially and politically irrelevant "zombie institutions." The article asks whether such degeneration has set in for APA and planning. Finally, it is concluded that more debate about APA's ethics and actions is needed for improving planning practice. Nine key questions are presented to constructively stimulate such debate.

**Highlights**: >A study of moral hypocrisy with the American Planning Association (APA). >"APA seriously breached its own ethics", according to *JAPA* editor. >APA is found to deny and divert evidence of malpractice and bad planning. >APA's hypocrisy is shown to place transparency and billions of dollars of citizens' money at risk. >Nine points for debate aimed at reducing hypocrisy with APA and improving ethics in planning.

**Keywords:** Planning, uncomfortable knowledge, moral hypocrisy, professional ethics, planning ethics, malpractice, the American Planning Association, *Journal of the American Planning Association*.

## UNCOMFORTABLE KNOWLEDGE

In organizational theory, uncomfortable knowledge is knowledge that is disagreeable or intolerable to an organization. Rayner (2011: 5-7) identifies four strategies in increasing order of sophistication for how organizations typically deal with uncomfortable knowledge:

1. *Denial* represents a refusal to acknowledge or engage with information.
2. *Dismissal* acknowledges that information exists, and may involve some minimal engagement up to the point of rejecting it as faulty or irrelevant.
3. *Diversion* involves the creation of an activity that distracts attention away from an uncomfortable issue.
4. *Displacement* occurs when an organization engages with an issue, but substitutes management of a representation of a problem for management of the problem itself.

What follows is a case study of how the American Planning Association manages uncomfortable knowledge about urban policy and planning. APA typically projects a "sunny, relentlessly positive" image of urban planning, in the words of an insider and informant for the present study. Thus APA presents itself as the promoter and guardian of the public interest, participation, transparency, openness, truth, social justice, and ethics in planning, as evidenced in APA's own publications. By taking a look behind the sunny image – at how APA handled a study containing uncomfortable knowledge about malpractice in planning – this article uncovers another side to APA, what Flyvbjerg (1996) and Yiftachel (1998) calls the "dark side" of urban policy and planning.

We will see that in this case APA's immediate reaction to uncomfortable knowledge was simple denial, i.e., strategy 1 above, which Rayner (2011: 7, 12) defines as the refusal or inability of organizations to acknowledge information, even when external bodies or individuals within the organization seek actively to bring it to collective attention. We will also see that, when pressed, APA escalated its strategy for dealing with uncomfortable knowledge from denial to diversion, diversion being defined as the organizational strategy of establishing a decoy activity that distracts attention from a subject or problem, thus trying to ensure that knowledge about it is not created or shared (Rayner 2011: 7, 12).

Finally, we will consider more productive strategies for dealing with uncomfortable knowledge than those employed by APA. In the case study, APA said it would not promote a study published by *JAPA*, which contained uncomfortable knowledge about malpractice and bad planning, to protect APA members' interests. Here it is argued that it is in members' interest to have malpractice critiqued and reduced, and that this can only happen by exposing malpractice, not by hiding it as APA does. Critique is historically a main driver of progress in human affairs. Professions, organizations, and societies that stifle critique tend to degenerate and become socially and politically irrelevant "zombie institutions" (Beck 1994: 40). This article raises the question of whether such degeneration and irrelevance applies to APA and urban planning and, if so, what can be done to counter it.[1]

## WHEN PLANNERS LIE WITH NUMBERS

The author first became aware of APA's approach to uncomfortable knowledge about urban policy and planning when he and his co-authors submitted an article to APA's flagship academic publication, the *Journal of the American Planning Association* (*JAPA*), called "Underestimating Costs in Public Works Projects: Error or Lie?" (Flyvbjerg, Holm, and Buhl 2002). The abstract of the article reads as follows:

> "This article presents results from the first statistically significant study of cost escalation in transportation infrastructure projects. Based on a sample of 258 transportation infrastructure projects worth US$90 billion and representing different project types, geographical regions, and historical periods, it is found with overwhelming statistical significance that the cost estimates used to decide whether such projects should be built are highly and systematically misleading. Underestimation cannot be explained by error and is best explained by strategic misrepresentation, that is, lying. The policy implications are clear: legislators, administrators, investors, media representatives, and members of the public who value honest numbers should not trust cost estimates and cost-benefit analyses produced by project promoters and their analysts."

The article shows that as a consequence of cost underestimation nine out of ten large public works projects have cost overruns. Cost overruns are large, even when measured in conservative terms, i.e., excluding inflation and using the final business case as base line. The study documents a cost overrun of 45 percent for rail projects, 34 percent for bridges and tunnels, and 20 percent for roads. Standard deviations are large, too, indicating risk to the second degree, i.e., risk of cost overrun *and* of the overrun being much larger than expected. Most interesting of all, overruns have been constant for the 70 years for which data are available, indicating that no improvements in estimating and managing costs have been made over time. Finally, the study raises serious concerns that cost underestimation appears to be deliberate in many cases.[2]

In sum, the study documents billions of dollars of cost overruns and waste that could, in large part, be eliminated by better policy and planning. It seemed obvious to the authors that the results must therefore be effectively communicated to policy makers and planners, which is why we submitted the paper to *JAPA*, following and expanding upon a tradition pioneered by Martin

Wachs's (1989) classic article, "When Planners Lie with Numbers," and by Kain (1990), Pickrell (1992), and Wachs (1990). We aimed to give this area of study more weight by contributing the largest and best dataset seen so far, making statistical analyses of cost underestimation and overrun possible for the first time and allowing firmer conclusions than previously about forecasting outcomes and forecasting behavior.

*JAPA* accepted the study for publication and agreed it documents a major issue in urban policy and planning that needs public attention. In the months leading up to publication, *JAPA* therefore contacted APA to enlist professional support for disseminating the study to media. Contacting APA was *JAPA*'s idea, but we agreed that seeking professional support for media contacts made sense. As it turned out, the person we liaised with at APA initially found the study "very newsworthy" and developed a comprehensive media strategy, including a press conference about the study; exclusives with *The New York Times*, *The Sunday Times* of London, and a major US television network; coverage by The Associated Press and Reuters; and finally regional press releases for media in the 8-12 largest US metropolitan areas. We agreed that APA would manage the media strategy in the US and we, with our university's press office, would manage it in the rest of the world

However, after a few weeks – while we were busy collaborating with APA on the press releases – our APA liaison suddenly wrote to let us know he needed to discuss the media strategy internally at APA. The effort had evidently grown to a size where he clearly felt uncomfortable by continuing it without involving top management. Soon after, he returned with the following discouraging message:

> "After discussion and consideration, [APA] does not want me to do anything in terms of promoting, whether as an exclusive story or on a regional basis, the results of your infrastructure cost underestimation study as appearing in the Summer issue of *JAPA* ...
I'm sorry I can't be of further help."

This was a complete turnaround in both tone and content, compared with APA's previous enthusiasm for the project and our extensive correspondence and the comprehensive publicity strategy developed by APA. Our joint work, including on the press releases, was stopped unilaterally by APA. A colleague got in touch with a former top-level official with APA, who had good connections within the organization, to find out what was going on. This person said that he believed "it was the 'higher ups' [in APA] who put the 'kabosh' [sic] on the press release[s]," but also that "the APA staff was [still] anxious to promote and publicize your work."

## DENIAL

When asked what the reasons were for APA's sudden change of heart, our APA liaison explained it was, first, fear that "the media will cast this story negatively and planners will be among the guilty" and, second, that "some of the study's finding[s] ... could work against us [APA and its members]." Our APA liaison further explained that he felt responsible for what he called APA's "180 degree switch" and that he found the study both exciting and newsworthy, exactly what journalists and editors are looking for. With a background in journalism, he had reacted to the study with the gut instinct of a journalist, seeing the potential for maximum media coverage, only to learn that APA would prefer no coverage at all:

> "I readily admit that I became excited about your research findings and the fact that *JAPA* is publishing a very newsworthy story. My background is journalism, and as a former daily and weekly newspaper reporter and editor ... I started to think about a publicity strategy for your article first and the possible negative implications for the planning profession second. In terms of my job at APA, I need to think about the latter."

Our contact at *JAPA* similarly explained:

"[APA] is worried that the spin will be that planners are liars, even though this is not what you say in your article, and [APA] is trying to protect the profession from that bad press."

Our APA liaison further pointed out that, "Given APA's concerns, I would encourage taking a much more restrained approach in the press release" than we had done in the release we had been collaborating on so far. Finally, our APA liaison said he had been asked by APA to draft such a more subdued release covering the study. This was bad news for the authors, because it meant that a powerful organization like APA would now write its own press release about our study. We would have no control over the APA release, which we understood would not be designed to emphasize the newsworthiness of the study, but would instead gloss it over in order to protect APA member interests. I asked our APA liaison to kindly let me comment on APA's press release before it was finalized. He agreed to this and said he would send me a copy for comment, but he didn't do so. Our contact with APA quickly died out as we made the move from partners to opponents, from collaboration to conflict. Commenting on this paper, our APA liaison now explains APA’s approach by saying “It was clear to us the article had high news value and, after discussing this point internally and with you – and learning of your own extensive efforts to obtain publicity – we concluded that the article did not need the APA’s promotion and publicity to generate news coverage.” However, this is not consistent with what we were told at the time, nor with the fact that our and our university's efforts to obtain publicity for the article in the US did not begin until *after* APA dropped its media strategy for the article and only because they dropped it.

At first, *JAPA* was on our side. *JAPA*'s management decided that *JAPA* would assert their independence of APA by single-handedly facilitating media coverage of the study in the US, no matter what APA said. Our contact at *JAPA* sent us the following message:

> "I was very disappointed to see [APA's representative, name deleted] latest note to you saying APA does not want him to work on PR for *JAPA* articles. I think our editors will be discussing this policy with the APA. Regarding your article, however, I'm prepared to do what I can to facilitate media coverage in the US."

But despite his good will, or perhaps because of it, within days our *JAPA* contact clearly came under pressure to retract his decision, now explaining:

> "I'm afraid I spoke out of turn in offering to pitch a story on your article to the US media ... APA has particular political reasons not to spotlight your findings. [We, at *JAPA*] discussed this ... [and]agreed that it would not be right for me to do this in light of APA's stance. I am therefore afraid I will have to retract my offer ... I apologize for this about face. I personally hope your findings get a great deal of media attention."

It was instructive to see how APA was able to produce one "180 degree switch" and "about face" after another. But power defines rationality, as argued in Flyvbjerg (1998: 227-34). This was power at work, making individuals do what they did not want to, against their basic beliefs. The personal greeting from our *JAPA* liaison that he hoped the study would get a "great deal of media attention," despite APA's determination not to promote the paper, testifies to this. In effect, APA's

"particular political reasons" seemed to be dictating the actions of a supposedly independent academic journal, the world-leader in its field, and thus violating rule number one in academic publishing: that such publishing must be thoroughly independent from outside interests.

Instead of APA openly acknowledging the implications of the *JAPA* study, that

- planners are doing an exceptionally poor job at costing major public works projects, sometimes perhaps intentionally,
- this results in large-scale waste of public money and violations of basic principles of democracy, and,
- APA, as the main professional body for planners, has a responsibility to help rectify this situation,

APA had decided to gloss this over and suppress publicity of the unpleasant facts. APA's idea of protecting member interests was apparently to deny and undermine malpractice concerns in planning, thus defending the status quo. Our idea was to improve planning by problematizing malpractice and demonstrating how it may be avoided.

To APA, the study on cost overrun was a clear instance of Rayner's (2011: 5-7) "uncomfortable knowledge," described above, i.e., knowledge that is disagreeable or intolerable to an organization. Of the four strategies for how organizations typically deal with uncomfortable knowledge – denial, dismissal, diversion, and displacement – here we see denial at work: APA refused to acknowledge the concerns about planners' apparent misrepresentation of costs in large public works projects, even when bodies like *JAPA* and individuals like the authors of the study were actively seeking to bring this knowledge to the collective attention of APA and its membership. When we pressed further to get the information heard, APA ramped up from denial to the more sophisticated strategy of diversion, which consists in establishing a decoy activity that attempts to distract attention from an issue, thus trying to ensure that knowledge about it is not created or shared (Rayner 2011: 7, 12). Let us see how APA did this.

## DIVERSION

Presumably in order to control information about the *JAPA* study, APA decided they would try to "track and monitor all reporter requests for copies [of the study]." To achieve this, APA now instructed *JAPA* to not get involved with media contact at all and instead "direct reporter inquiries to [APA] and we [APA] will follow-up [sic] ... We are doing this so that if we want to talk to a reporter about the study, we don't give him or her a copy of the study until we've had a chance to talk with them." According to APA's plan, neither *JAPA* nor the authors would have a say in what the media were told by APA about the study.

*JAPA* accepted APA's machinations, we did not. It was less than three weeks to publication when APA's attempt at controlling public exposure of the study hit us out of the blue, and we quickly decided to try and beat APA to the media. First, we personally pitched our story and offered exclusives to the editors of *The New York Times* and *The Sunday Times* of London, which they both accepted.[3] We figured that if two prominent newspapers like these would cover the study, then the rest of the media would follow. This proved to be correct. We realized we did not need PR people to communicate with media. In fact, editors and journalists seemed to prefer to deal with us directly. Second, while *The New York Times* and *The Sunday Times* of London were printing their exclusives, we emailed copies of the study and a press release to every other main media in the English-speaking world, including the US. Many of these media covered the study.

It must have taken APA a few days to realize they had little sway over who received the study and how it was covered. APA then ramped up its strategy from denial to diversion, i.e., APA now actively tried to distract attention away from the study and the uncomfortable issues it raised. APA did this by posting a public comment about the study on its website, in effect a disclaimer that tried to paper-over and downplay the study's findings:[4]

> "While we [APA] can appreciate the technical skill and data collection capability employed in the article, as well as the policy questions that are raised, readers of this technical article should bring their own context to the discussion ... Planners ought to be concerned not only with the costs of public works ... There is a danger in being overly analytical about the details of processes that are ultimately not empirical so much as they are democratic and political – a behavior that Alfred North Whitehead called, aptly, 'misplaced concreteness.' " (American Planning Association 2002).

APA's public comment was written by our APA liaison. It is interesting to note that in the APA comment he now tried to frame the study as "technical," by twice labeling it as such, and as being "overly analytical," thus undoubtedly encouraging a certain reserve with readers against the study. This was at complete odds with his own views about the study expressed to us previously, when he had described it as "very newsworthy" and himself as "excited about your research findings."

It is also interesting to note APA's artificial juxtaposition in the comment of "empirical" on one side and "democratic and political" on the other, as if democratic and political issues cannot be studied empirically, which is plain wrong. Finally, things get even more obfuscated with the deployment of Whitehead, the heavyweight mathematician, logician, mentor and collaborator of Bertrand Russell, and someone who would recognize a technical argument if he saw one. But unfortunately APA gets Whitehead wrong. The fallacy of misplaced concreteness is about treating abstractions as realities. It is not about being overly analytical about details or being empirical at all, but quite the opposite. APA's use of Whitehead appears to be a botched attempt at oppressive deployment of symbolic capital rather than an actual rational argument. Fortunately, the media did not buy APA's interpretation of the study but covered it instead like APA had originally intended for them to do, i.e., as newsworthy and exciting.

In its public comment on the *JAPA* study APA also boosted its own high ethics in the manner that has become a trademark of the organization, taking a normative stand for participation, transparency, the public interest, and good information:

> "This [the technical character of the study, etc.] doesn't mean planners shouldn't get involved. Of course planners have an ethical obligation to promote sound project planning and the greatest possible degree of public involvement, transparency, and the understanding of the options available ... [E]ven if [planners] are not necessarily able to effect as much control over the process as would be useful to the public's interest, [planning] still requires their strong and active presence and voice for good information and an open and ethical process" (American Planning Association 2002).

It is interesting to note that APA's comment, which stresses the virtues of transparency, ethics, etc., itself seems to be a deliberate and direct attempt by APA to block transparency, information, and openness for an important but uncomfortable planning issue, namely cost overruns of billions of dollars on major planning projects. APA's words and actions are diametrically opposed on this point, showing a double standard that is used by APA because it feared the *JAPA* study might present planners in a bad light. APA's comment and its double standard can only be seen as an attempt at managing uncomfortable knowledge by the "diversion strategy" described above, where the comment and its obfuscating and misleading content functions as a decoy activity that aims at distracting attention from the issue at hand – here planners who appear to systematically misrepresent the costs of projects – thus hampering the spread of knowledge about this issue.

## IS *JAPA* TRULY INDEPENDENT?

The perhaps most surprising thing about APA's attempts at diverting attention away from the uncomfortable issues raised by the cost study is the fact that the *JAPA* editors let APA do its maneuvering without weighing in against it. Scholars generally believe that scholarly values must be upheld at all times for scholarly work. The independence of journal editors and their journals is one such core value. Publication of the study was the *JAPA* editors' responsibility, because they had approved it. Although *JAPA* did in fact publish the study, APA, their parent organization and sponsor, refused to let *JAPA* promote the study and actively tried to undermine it in public, as we saw with APA's public comment above. Several of our US colleagues saw this as APA "second guessing" the *JAPA* editors. One colleague, who served on AICP's (American Institute of Certified Planners) Ethics Committee for several years before the *JAPA* study was published and in other official roles later, put it like this:

> "I feel that if there is one principle that is strong and must be honored continuously, it is the independence of the editorial review process from the elected members of the APA Board and the AICP Commission.[5] So I would also suggest that it might be appropriate for the editor [of *JAPA*] to comment [on APA's actions regarding the study] ... on the propriety of them [APA] second guessing the editorial process."

This commentator later added the following viewpoint about the relationship between *JAPA* and APA, placing responsibility for the unsatisfactory treatment of our article not only with APA but also with *JAPA*:

> "The relationship between the editor of the *JAPA* and the APA staff is puzzling to me. A good editor could have been more aggressive, so I am not sure the problem resided mostly with the APA staff."

The lack of reaction on *JAPA*'s part indicated that *JAPA* might not have the true independence from outside forces, including from APA, its sponsor, that an academic journal must have to be fully credible. This is not to say APA interfered with *JAPA*'s decisions on publication, which are made by its editor based on double-blind peer review. But in this case APA interfered with the public exposure of a *JAPA* study, trying to control and "spin" it to protect APA's boosterish image of planning. With no comment or other reaction forthcoming from *JAPA*, our US colleagues suggested we ask APA for permission to write a comment ourselves on how we saw APA's actions. A few days after APA published its comment on the study I therefore asked APA for permission to write a rejoinder. But again APA practiced its denial response. I received no answer, including to a later reminder.

Two weeks after APA's public comment about our study and almost a month after its publication in *JAPA*, our contact at the journal reported that there continued to be a "flurry of requests for the article and the subsequent news stories" and that he had received a request from the Sierra Club for permission to reprint the article as part of a protest against a project. This would be the first of many similar requests from groups opposing major projects, who would use the research to resist specific projects. Finally, *JAPA* had received a letter to the editors about the article, to which my co-authors and I duly responded (Flyvbjerg, Holm, and Buhl, 2003). The report from *JAPA* confirmed what we already knew: Despite APA's attempts to deny and divert attention away from the study, we had succeeded in effectively communicating its results to the public sphere, as we had set out to do, for the results to inform policy and practice. This would eventually lead to policy changes in a number of countries on how major public works projects are planned and managed (Flyvbjerg 2012: 176) and to an endorsement of our recommendations from the godfather of behavioral economics and Nobel Prize winner in economics 2002, Daniel Kahneman, as "the single most important piece of advice [that exists] regarding how to increase accuracy in forecasting" (Kahneman, 2011: 251). The policy changes and endorsement would most likely not have happened if APA had had its way with its attempt to suppress publicity of the 2002 *JAPA* study. Later, as we published more on better forecasting theory and methodology, including in *JAPA* (Flyvbjerg, Holm, and Buhl, 2005), APA too would find reason to endorse our thinking (American Planning Association, 2005b). APA clearly found it easier to embrace our constructive work than our earlier problematizations, although the latter was a precondition for the former.

## THE *REAL* ETHICS OF APA

In its efforts to deny, spin, and divert attention from the *JAPA* study on cost overrun, we consider that APA violated its own Code of Ethics on at least six counts. First, the APA/AICP Code of Ethics specifically states that a planner's primary obligation is to serve the public interest (American Planning Association 2005a).[6] It would clearly serve the public interest that the public got to know about the billion-dollar cost overruns, misinformation, and waste of public resources that are typical of major public works projects and are documented by the *JAPA* study. By trying to suppress and divert publicity for the study, APA was actively working against the public interest and thus against the "planner's primary obligation."

Second, the APA/AICP Code of Ethics stipulates that a planner must strive to provide full, clear, and accurate information on planning issues to citizens. The *JAPA* study strives to do exactly that by publishing the largest and best dataset and set of analyses that exist on cost underestimation and overrun in major planning projects. With its attempt to suppress publicity and divert attention from the study, APA was actively attempting to hinder such information in reaching citizens and was thus violating its own Code of Ethics.

Third, a planner must accurately represent the qualifications, views, and findings of colleagues, according to the Code of Ethics. With APA's public comment on the *JAPA* study, posted prominently on the APA website, APA deliberately tried to put its own spin on the findings of the study, as shown above, and not just in our view, but in the view of colleagues, members of APA, and people who held or had held office with APA. APA thus violated its own Code of Ethics on this point, too.

Fourth, the Code of Ethics states that a planner must share the results of experience and research, which contribute to the body of planning knowledge. With the *JAPA* study, we were trying to do just this. With their diversion strategy and tactics, described above, and their attempts at suppressing media coverage of the study, APA management was trying to hinder such sharing of results, and thus worked against the APA/AICP Code of Ethics.

Fifth, a planner must not commit a deliberately wrongful act, which reflects adversely on the planner's professional fitness, says the Code of Ethics. Here, the actions of APA regarding the *JAPA* study were, in our view, wrongful and deliberate, despite the Code saying they "must not" engage in this type of behavior. As such their actions reflect adversely on APA's professional fitness.

Finally, the APA/AICP Code of Ethics specifies that a planner must systematically and critically analyze ethical issues in the practice of planning. With the *JAPA* study we attempted to do this. However, with its papering over of the results and its efforts at suppressing publicity, APA systematically tried to suppress knowledge of our critical analysis and to limit its impacts on the public and on the planning profession, breaching its own Code of Ethics one more time.

That's a lot of "musts" disregarded by the very body that stipulates that the "musts" must be observed. It is a matter of serious concern that the AICP Ethics Officer, who, in collaboration with the AICP Ethics Committee, is responsible for upholding the Code of Ethics is not independent but is also APA's Chief Executive Officer, the same person today as when the *JAPA* study was published. A former member of the AICP Ethics Committee who commented on the present study observed about this that the lack of an independent Ethics Officer within the structure of APA is a fundamental problem and that this "set the stage" for APA's dubious treatment of the *JAPA* study. In 2001, an independent counsel, who had been asked by AICP to assess its Code of Ethics, recommended the establishment of an independent Ethics Officer. However, APA's CEO cum Ethics Officer was against the proposal and successfully blocked it, according to the Committee member. This has left APA with its questionable practice of conflating in one and the same person the roles of CEO and Ethics Officer. The consequence is a lack of arm's length principle in APA's ethics.

## NINE QUESTIONS FOR A DEBATE ABOUT APA'S ETHICS

Upon reading the above analysis, an editor at *JAPA* commented, "I agree that APA seriously breached its own ethics" and that it "is a good summary of some serious APA shenanigans." A former APA president similarly called the study a "sad but true comment on the way APA protects what it perceives as its turf." Then the ex-president made a remark that surprised me:

> "I agree with your interpretation of events, except I think *you may be a bit soft on APA and its code of ethics*" (emphasis added).

I was surprised, because I thought that the analysis of APA's ethics above might be seen as too hard, despite all efforts at being as balanced and evidence-based as possible. But the former APA president explained that to an insider like himself there are more basic issues at stake with APA's ethics than what is uncovered here. To him the real problem is that after having been involved with APA for several decades he cannot recall a single example of a planner being expelled from APA for ethical violations[7] According to the former president this is not because planners are uniformly well-behaved, but because APA is in denial about the possibility of bad planning and malpractice and tends to see all planning as good.[8] That is the real problem with APA, according to the former president, and in relation to this larger problem the analysis above seems to him just the tip of the proverbial iceberg and therefore soft on APA. However, given the fundamental nature of the issues at stake for planning as a profession and the principled debates needed to deal with them, I would rather err on the side of softness, because being too hard runs the risk of the critique being rejected as exaggerated and thus having no effect.[9]

Nevertheless, the former APA president is right that APA appears to be in denial about bad planning and potential malpractice and that this is a fundamental problem. To underestimate billions of dollars of costs in planned projects, and thereby cause billions of dollars of cost overruns, is bad planning by any definition of the term. And if underestimation is deliberate then it is lying, which is unethical by any code of conduct and unlawful in many jurisdictions for this type of issue (Bok 1979, Cliffe et al. 2000, Wachs 1989).[10] Such bad planning and unethical behavior place citizens, taxpayers, investors, and other stakeholders in public projects at a risk that is deliberately hidden from them. APA helps hide the risk to the extent that APA suppresses publicity, and manipulates the spread, of knowledge about it, as we saw APA do above. This demonstrates a lack of concern over bad planning and malpractice which must result in a lack of accountability, without which a profession cannot make proper progress. To not take malpractice concerns and accountability seriously is to not take the profession of planning seriously. We therefore arrive at the paradoxical conclusion that APA is not taking the profession of planning seriously and is holding back its development. APA's management say they protect member interests, but they do quite the opposite. It is in members' interest to have malpractice problematized and reduced, and this can only happen by exposing malpractice, not by denying or diverting attention from it as APA did in the case of the *JAPA* study described above.

Lord Acton famously said that power tends to corrupt, and absolute power corrupts absolutely. APA is the biggest and most powerful professional organization for planners in the world. Backed by this kind of authority, it is cause for worry if APA violates its own Code of Ethics and sweeps under the carpet concerns about bad planning and malpractice, as we saw above, so as to not upset the boosterish image of planning that APA is at pains to uphold. Such moral hypocrisy should give planners, planning academics, and planning students pause to think. We need to ponder and debate the real ethics of our profession in order to improve upon it. The present article was written to constructively contribute to this type of debate and we end by raising the following nine questions about APA's ethics and actions for serious discussion and dialogue, as opposed to polemics:

1. How can the planning profession improve and make progress if critique and self-critique of planning and planners are not addressed by our biggest and most powerful professional organization, APA?

2. How can planners become better if they are always already portrayed as good by APA – in the boosterish image of planning that APA paints – and if APA deliberately turns a blind eye on and glosses over planners doing bad planning, i.e., planning that violates basic norms of democracy, efficiency, and equity?

3. Is APA so deeply in denial about bad planning and malpractice that APA is part of the problem instead of the solution?

4. Is it really true, as suggested by a former APA president, that APA practically never excludes members due to violations of its Code of Ethics? How do the numbers and types of reported and sanctioned code violations for planners compare with those of other professions, including medicine, law, and accounting? How have the numbers developed over time?

5. What should be done and by whom about APA's moral hypocrisy, documented above, where APA seems to disregard its own Code of Ethics, will undermine critical reviews, and will block transparency in order to deny and divert information about and critique of bad planning and malpractice?

6. Is there a fundamental conflict between being an advocacy body and being the guardian of professional ethics? Between booster and regulator? Should the two functions be separated? Does APA need a fundamental rethink of its role in those terms?

7. Does APA/AICP need an independent Ethics Officer, as proposed by an independent counsel in 2001? Does the fact that APA's Ethics Officer is also APA's CEO lead to a conflict of interests and lack of arm's length principle? Do other professional organizations conflate these roles like APA does?

8. What are the consequences of a boosterish professional organization like APA being linked to its field's major academic publication, *JAPA*? Is *JAPA* truly independent of APA? Is there a conflict of interest? If yes, what should be done about it?

9. Critique is historically a main driver of progress. Professions, organizations, and societies that stifle critique tend to degenerate and become socially and politically irrelevant zombie

> institutions. Is APA and planning in danger of such degeneration and irrelevance? Has it already set in? What should be done, if anything?

By debating these and similar questions, perhaps we can begin to see APA and planning for what they are, warts and all, and then work to improve them. Instead of being stuck with the relentlessly positive picture-book image of planning that APA and many planners have been painting for too long to the detriment of progress for the profession. Planning is too important to be treated like that.

If planners choose to *not* engage in this debate, the risk is that ever more members of APA and AICP will follow the example of this former AICP member, who used to be seriously engaged and hold high office with the organization, but who wrote me after having read the analysis above:

> "I need to tell you that I have let my membership in AICP lapse. This may be a surprise or perhaps it is not. I am no longer entitled to put FAICP after my name. I think that the AICP is an organization founded upon principles of self promotion and self interest among planners [as opposed to the public interest stressed in the AICP Code of Ethics] ... In order to be a member, a person has to engage in 'continuing education,' and to amass continuing education credits by enrolling in courses, attending conferences, and so forth. They state that the purpose is to insure high quality and professional job performance by professional planners. The cost to attend the national AICP conference is [high], and I found the sessions to be boring and shallow. I interpreted the fees charged as being in the interest of the organization which is run like a business ... I made a personal choice to stop accumulating 'credit' by attending specific events that offered credit unless I happened to want to attend them. So, I was eventually informed that my membership in AICP would 'lapse' this year unless I signed up for a certain number of units of credit. I informed AICP that I preferred to

> allow my membership to lapse ... This reflects my own assessment of their professional status and standing."

## SUMMARY AND CONCLUSIONS

With a point of departure in the concept "uncomfortable knowledge," this article presents a case study of how the American Planning Association deals with such knowledge, in the form of knowledge about potential malpractice by planners and bad planning. APA is found to employ two well-known strategies for dealing with uncomfortable knowledge: denial and diversion. APA actively suppress publicity of and undermine information about malpractice concerns and bad planning in order to sustain a boosterish image of planning and planners. In the process, APA appears to have disregarded and violated its own Code of Ethics on multiple counts. In so doing, APA is potentially placing principles of transparency and democracy at risk, together with billions of dollars of citizens' money. APA justified its actions as necessary to protect APA members' interests, i.e., prevent planning and planners from being presented to the public in a bad light. The article concludes, however, that it is in members' interest to have malpractice critiqued and reduced, and that this best happens by exposing and addressing malpractice concerns, not by denying or diverting attention from them as APA did in this case. Critique is historically a main driver of progress. Professions, organizations, and societies that stifle critique tend to degenerate and become socially and politically irrelevant. The article asks whether such degeneration has set in for APA and planning. Finally, it is concluded that more debate about APA's ethics and actions is needed for improving planning practice. Nine key questions are presented for stimulating such debate.

# NOTES

[1] The research for this paper was carried out during and immediately after the events described. Main sources quoted in the article have been given opportunity to comment on earlier drafts and their comments have been incorporated in this final version, where relevant. Publication of the research has been delayed in order to protect informants, several of whom acted like O'Leary's (2010: 8) "guerrilla employees." Many of these have now moved on to other positions. For further protection, all sources have been anonymized. The author knows the names of all sources quoted in the study. Documentation of all quotes and other data are on file in the author's archives. Finally, publication was also delayed in order to emphasize that the study is about matters of principle – malpractice and professional ethics in planning – and not about individuals and individual actions.

[2] A former APA president provides the following explanation of the reasons why planners misinform about costs and why APA does nothing about it: "I believe planners and consultants in general deliberately underestimate project costs because their political bosses or clients want the projects. Sometimes, to tell the truth is to risk your job or your contracts or the next contract. I think the professional organizations [like APA] are likely to simply point to their code of ethics and let it go at that [i.e., enforcing no sanctions for those who violate the code]. For example, Article B 11 in the AICP [American Institute of Certified Planners] code says: 'Planning process participants should not misrepresent facts or distort information for the purpose of achieving a desired outcome', [but I have not seen it enforced]." (Personal communication, author's archives). The AICP is APA's professional institute, providing certification of professional planners, ethical guidelines, professional development, and more.

[3] Printed on July 11 and 7, 2002, respectively.

[4] A colleague, who had previously held a high-ranking position within APA, and who commented to us on the unfolding of events, sent us this observation about APA's comment: "[APA] seems to want to paper-over your scholarly findings and downplay the importance of your article ... I think this is important."

[5] The AICP Commission is the governing body of AICP.

[6] This is the APA/AICP Code of Ethics that applied in 2002, when our study came out (American Planning Association 1991). The Code was revised in 2005, but still contains similar rules of conduct (American Planning Association 2005a).

[7] The author has checked this statement with other APA insiders and with APA's CEO and found that exclusions of members of APA/AICP for ethical violations seem very rare; like the former APA president no one appears to be able

to remember specific instances. According to the CEO, however, "[t]here have been individuals whose membership has been revoked as a result of ethics proceedings" (personal communication, author's archives).

[8] The former APA president explains: "I've been involved with APA for about 40 years having served as [several top-level positions deleted here for reasons of anonymity] and I don't recall more than a handful of planners being brought up on ethical charges and no one being drummed out for ethical violations. APA believes all planning is good; note the slogan 'Making Great Communities Happen' and the sunny, relentlessly positive contents of its *Planning Magazine*."

[9] A former member of the AICP Ethics Committee similarly commented that the article "doesn't go far enough" in placing responsibility with specific people within APA for APA's apparent lack of adherence to its own Code of Ethics. For full disclosure, it should be mentioned that the study did uncover material that is substantially more critical of specific individuals than what is revealed here. A surprising number of respondents, who worked for APA as officials or employees, were remarkably critical and even resentful of the organization and how it is managed. Such resentment may be an interesting topic for further study, but is not the issue here. Here focus is on APA's ethics and knowledge management.

[10] It is worth noting that if, instead of public planners, it were employees of private corporations who deliberately underestimated costs like documented in the studies mentioned here, they might be committing a criminal offense under the Sarbanes-Oxley Act (2002). The Sarbanes-Oxley Act (Section 802[a], 18 U.S.C. § 1519) states that whoever knowingly alters, destroys, mutilates, conceals, covers up, falsifies, or makes a false entry in any record, document, or tangible object with the intent to impede, obstruct, or influence the investigation or proper administration of any matter within the jurisdiction of any department or agency of the United States or any case filed under title 11, or in relation to or contemplation of any such matter or case, shall be fined, imprisoned not more than 20 years, or both. This is not to suggest that planners engage in criminal activity, but to consider that similar behavior in another context might be criminal.